**Structure and Non-Equilibrium Heat-Transfer of a Physisorbed Molecular Layer on Graphene**

*Bareld Wit\*, Ole Bunjes, Martin Wenderoth and Claus Ropers\**

B. Wit, O. Bunjes. M. Wenderoth, C. Ropers
IV. Physikalisches Institut, Georg-August-Universität Göttingen, 37077, Germany
E-mail: bareld.wit@uni-goettingen.de, claus.ropers@uni-goettingen.de

The structure of a physisorbed sub-monolayer of 1,2-*bis*(4-pyridyl)ethylene (bpe) on epitaxial graphene is investigated by Low-Energy Electron Diffraction and Scanning Tunneling Microscopy. Additionally, non-equilibrium heat-transfer between bpe and the surface is studied by Ultrafast Low-Energy Electron Diffraction. Bpe arranges in an oblique unit cell which is not commensurate with the substrate. Six different rotational and/or mirror domains, in which the molecular unit cell is rotated by 28±0.1° with respect to the graphene surface, are identified. The molecules are weakly physisorbed, as evidenced by the fact that they readily desorb at room temperature. At liquid nitrogen temperature, however, the layers are stable and time-resolved experiments can be performed. The temperature changes of the molecules and the surface can be measured independently through the Debye-Waller factor of their individual diffraction features. Thus, the heat flow between bpe and the surface can be monitored on a picosecond timescale. The time-resolved measurements, in combination with model simulations, show the existence of three relevant thermal barriers between the different layers. The thermal boundary resistance between the molecular layer and graphene was found to be $2\pm1\cdot10^{-8}$ K m$^2$ W$^{-1}$.

**1. Introduction**

The properties of ultra-thin molecular layers on surfaces are governed by the delicate interplay between adsorbate-adsorbate and adsorbate-surface interactions.[1] Hence, understanding these interactions is of paramount importance for the development of applications involving surface-bound functional molecules, such as molecular (opto-)electronics, molecular machines, catalysts, smart coatings, and chemical sensors.[2] Equilibrium measurements can be used to identify, for instance, the adsorption geometry, long-range order and defects, covalent and non-covalent molecule-molecule interactions, or possible phase transitions.[3] This requires techniques which are exceptionally surface sensitive and simultaneously capable of resolving detailed structural features. Commonly used techniques which satisfy both requirements include Scanning Tunneling Microscopy (STM) and Low-Energy Electron Diffraction (LEED).[4]

However, many elementary processes, such as vibrational relaxation, molecular motion, and bond formation/breaking, occur on femto- to nanosecond timescales.[5] Therefore, experimental techniques with high surface sensitivity and high temporal resolution are desirable. Specific ultrafast (non-)linear spectroscopic methods are able to track changes at surfaces,[6] but the development of techniques that can additionally resolve structural details to a high degree has proven to be exceedingly challenging. Nonetheless, time-resolved variants of Reflection High-Energy Electron Diffraction[7] and STM[8] are now available. Moreover, Ultrafast LEED (ULEED) has recently been developed in our group[9,10] to enable the investigation of dynamical phenomena on surfaces with a temporal resolution down to a picosecond.[11]

The combination of high temporal resolution and surface sensitivity is also advantageous for the investigation of thermal properties at the nanoscale.[12] Specifically, the Thermal Boundary Resistance (TBR or $R_K$ in K m$^2$ W$^{-1}$)[13] of interfaces becomes increasingly important as the characteristic dimensions of the system approaches the so-called Kapitza length $L_K = R_K\, k$, where $k$ is the thermal conductivity in W m$^{-1}$ K$^{-1}$ of the bulk material.[14,15]

The TBR receives significant attention owing to the importance of heat dissipation control in semiconductor microelectronic devices.[15,16] Likewise, heat transport across molecule/solid interfaces is of interest for future applications including molecular electronics and nanothermoelectrics.[17,18] One powerful approach to investigate the TBR is to measure the cooling of a thin material layer on a substrate following pulsed-laser heating.[12] Experimental techniques that use, for example, transient changes in reflectivity,[19] diffraction[20,21] or sum frequency generation[22] as well as theoretical methods[23,24] have been used to investigate TBRs of different systems.

In this work, sub-Monolayers (ML) of 1,2-*bis*(4-pyridyl)ethylene (bpe, see **Figure 1** A) on epitaxial graphene are investigated. In section 2.1, a thorough characterization of the structure of bpe on graphene by STM and LEED is presented. The molecule orders in small domains of an oblique unit cell, which is not commensurate with the surface. Section 2.2 details time-resolved measurements and model simulations of non-equilibrium heat transfer between bpe and graphene. ULEED is employed to probe the heat transport between the different layers of the bpe/graphene system after heating with a femtosecond laser pulse. We find that the overall heat dissipation takes place on the order of a few hundred picoseconds. However, the observation of a faster component in the temperature decrease of the bpe molecules, in combination with simulation, reveals the existence of a thermal barrier between bpe and graphene. Finally, the magnitude of the observed barrier between bpe and graphene is put into the context of existing literature to show that the thermal coupling between the graphene and the molecules is relatively strong, despite the weak Van der Waals interactions.

## 2. Results

### 2.1. Structural Characterization

*2.1.1. Clean surface*

Before discussing the bpe-covered surface, it is useful to briefly address the structure of the clean sample. The substrate consists of three distinct parts: 6H-Silicon Carbide (SiC), a Graphitic Buffer Layer (BL) and Few-Layer Graphene (FLG), as shown in Figure 1 B. BL is covalently bound to SiC and is (6√3 x 6√3)R30° reconstructed. The FLG layers are grown epitaxially on top of the buffer layer in a single domain rotated by 30° with respect to the bulk SiC.[25] High-resolution STM images of the graphene surface (see **Figure S1**) show a short-range hexagonal structure corresponding to the graphene lattice. A long-range contrast modulation, where areas much larger than a single graphene unit cell appear either brighter or darker than average, can be attributed to the influence of the buffer layer.[26] In LEED patterns of the clean surface recorded at 145 eV electron energy, spots of all three components can be identified (see **Figure S2** A). Only the first-order reflections of SiC and FLG can be seen, but a large number of spots originating from BL are present. At 60 eV, only buffer layer reflections are within the detector range (see Figure S2 C). Note that only a specific selection of the BL spots can be observed (see Figures S2 B&D). The relative intensities of SiC, BL and FLG spots in the LEED pattern of the clean surface can be used to show that FLG consists of 1-2 graphene layers depending on the location.[27]

*2.1.2. STM*

Now, we discuss the STM characterization of the bpe-covered graphene surface. Large-scale overview images, such as the one in Figure 1 C, show bright areas, corresponding to ordered molecular islands, and dark areas, representing the bare graphene surface. Some long-range contrast modulation in the bare areas can be seen, similar to that of the clean surface. The height of the islands corresponds to a single layer of bpe molecules. No evidence for the formation of multilayers has been found. Even at this large scale, a clear periodic contrast change due to the molecular arrangement can be seen in the bpe-covered areas. However, this pattern has a different appearance in different areas due to the existence of different mirror and rotational domains. These domains are typically on the order of 10 nm in diameter and single bpe islands are often an amalgamation of multiple domains.

In STM images with sub-molecular resolution, such as Figure 1 D, individual molecules can unambiguously be discerned. In Figure 1 F, a tiling of a spatially averaged unit cell is also shown in order to highlight the detailed features of the observed structure. Each molecule appears as a rectangular entity with four bright features at the corners and a stronger central bright feature. They can be attributed to the pyridyl moieties, with the nitrogen atoms appearing dark, and the ethylene bridge in-between, respectively. This shows that the molecules absorb in a flat-lying configuration with the pyridyl rings parallel to the surface. The molecules arrange in a herringbone pattern where the molecules with different orientations are also chiral mirror images. This is in accordance with the fact that bpe is pro-chiral. With this information, the unit cell can be identified: it belongs to the p2 plane group and has approximate lattice parameters $a = 1.61\pm0.02$ nm, $b = 1.35\pm0.02$ nm and $v = 37\pm2°$. A schematic unit cell structure is shown in Figure 1 E. Even though these values are extracted from STM images without an internal calibration, they agree well with the unit cell parameters found for bpe on recompressed exfoliated Highly-Ordered Pyrolytic Graphite (HOPG) by X-Ray Diffraction (XRD).[28]

*2.1.3. LEED*

We now discuss the characterization of the bpe islands with LEED. Upon deposition of bpe on epitaxial graphene, a rich LEED pattern can be observed, as shown in **Figure** 2 A. At the electron energy used, the first-order FLG and SiC reflections are located outside the detector range. Additionally, only the most intense spots of the $(6\sqrt{3} \times 6\sqrt{3})R30°$ reconstructed BL can still be seen (green circles in Figure 2 A). On the other hand, a large number of spots originating from the molecular islands are present. The overall six-fold symmetry of the molecular diffraction pattern in Figure 2 A is consistent with the existence of six rotational and mirror domains. The Full-Width-Half-Maximum (FWHM) of the diffraction spots, as shown in Figure 2 E, gives a lower limit of approximately 20 nm for the correlation length in the domains.

In order to index the bpe spots, unit cell parameters for this molecule on HOPG are used to simulate diffraction patterns.[28] While some exceptions exist,[29] it can be expected that bpe orders similarly on the two surfaces. The reciprocal unit cell of a single domain is pseudo-rectangular, as shown in Figure 2 B. However, different equivalent domains can be obtained by rotation or reflection, as indicated by the $C_6$ and σ operations in Figure 2 B. Overlaying the reciprocal unit cells of all six domains gives the pattern in Figure 2 C. This pattern reproduces the experimental data well, as illustrated by the two most prominent features highlighted in Figure 2 D. A quantitative comparison of the experimental diffraction pattern and the model, shows an agreement within ±3 % for reciprocal distances and ±1° for angles. The good agreement between experiment and model shows that bpe forms the same monolayer structure on FLG as on HOPG.[28] However, we can obtain additional information from the LEED images, as shown below.

Not all spots that are predicted can be observed; some are very weak or entirely absent. Notably, the {1 0} and {1 1} spots, which are predicted to lie within the first ring of observed spots, are missing at electron energies between 50 and 150 eV. The absence of these specific spots could be due to a structure factor where two scattering centers, located half a unit cell apart in the **a** direction, interfere strongly with each other. This would lead to a diffraction pattern reminiscent of that of a smaller lattice, as indicated by the dashed line in Figure 1 E. Closer inspection of the molecular arrangement reveals that a central double bond of a molecule is located at each corner of this smaller unit cell. While the two inequivalent molecules in the original unit cell have very different orientations with respect to the surface, their central double bond is rotated only slightly. This feature is coincidently also illustrated by the STM images in Figures 1 D & F, where the bright feature in the center of each molecule is aligned in approximately the same direction.

Due to the low symmetry of the bpe unit cell, no small commensurate structure can be formed. However, six rotational and mirror domains with a well-defined angle of 28.4±0.1° to the graphene high-symmetry directions are observed. This suggests that the structure is most likely a Higher Order Commensurate (HOC) structure rather than a true free-floating incommensurate layer.[30] While the two cases cannot be distinguished in STM or LEED directly, it is worth noting that no significant unit cell contraction was observed with LEED during cooling from Room Temperature (RT) to Liquid Nitrogen Temperature (LNT). This is further evidence that the layer is likely pinned and thus of HOC type. Furthermore, the formation of a HOC structure introduces additional strain in the bpe layer, which might explain the small size of the islands.

At RT, the same diffraction pattern can be observed, but with significantly reduced spot intensities. This shows that no structural phase transition occurs in the temperature range between RT and LNT. However, the strong attenuation of the intensities suggests that a shift in the equilibrium between islands and freely diffusing molecules takes place. Also, even though bpe was successfully deposited on the substrate at RT, it was found that it is not stable at this temperature in Ultra-High Vacuum (UHV); all diffraction spots associated with the molecules vanish within half an hour. This low re-evaporation temperature contrasts with the observation of a 2D structural phase transition for bpe sub-MLs on recompressed exfoliated HOPG at 414 K by XRD and Differential Scanning Calorimetry.[28] This discrepancy can be attributed to the different sample environments: a powder in low vacuum versus a surface in UHV.[31] At temperatures below 250 K, no significant desorption was observed, and after cooling to LNT, the molecular layer was stable for at least several days.

**2.2. Non-equilibrium heat transport**

*2.2.1. ULEED*

In order to investigate heat exchange between bpe and FLG and dissipation to the substrate, pump-probe measurements were performed at LNT. No phase transition could be induced by the pump laser, in agreement with the observations during cooling. However, LEED spot intensities can be used as a sensitive temperature probe: diffraction intensities are diminished with increasing vibrational disorder in the scattering centers, and thus with increasing temperature.[20] This relation is described by the temperature factor, or Debye-Waller factor.[32] The intensity change can directly be related to a temperature change, assuming that vibrations equilibrate sufficiently fast.[5] Additionally, the fraction of molecules detaching from islands should be insignificant; comparing the relative changes in spot intensities in time resolved measurements with those in static measurements during cooling shows that this is indeed the case.[33] Moreover, since in this system different diffraction spots can uniquely be assigned to bpe and the buffer layer, the temperatures of these layers can be detected independently. Note that the temperature changes in FLG cannot be measured concurrently with the other layers, due to its much larger reciprocal unit cell. Differences in heating or cooling rates obtained in a time-dependent measurement can provide valuable insight into the TBR between the materials in question.

In the ULEED experiments, we find that, indeed, the spot intensities are suppressed after the arrival of the optical excitation, by approximately 9 and 13 % for BL and bpe respectively. **Figure 3** A shows the temperature-time (*T-t*) curves for bpe and the buffer layer. The initial fast exponential rise of the temperature in both layers with a time constant of approximately 40 ps is at the temporal resolution limit of the present experiment. Thus, only limited information is extracted from this regime. However, a slower cooling process follows this fast heating. Fits to the bpe and BL cooling with double and single exponential functions, respectively, show that the characteristic time-scale for the overall cooling in both layers is in the order of $t_{BL} = 400\pm25$ ps $\approx t_{bpe,long} = 480\pm100$ ps. In addition, there also appears to be a small deviation from a single-exponential decay in the case of bpe: a faster initial cooling rate with a characteristic time scale of about $t_{bpe,short} \approx 100\pm50$ ps. The shaded area in Figure 3 A indicates the difference between the bi-exponential function and its slower component.

*2.2.2. Model*

In order to gain additional insight into the physics governing this transient increase in the temperature of bpe and BL, qualitative simulations of the heat dissipation have been performed. A schematic showing the different layers, the TBRs investigated and the initial temperatures is shown in Figure 3 B. In general, the bpe and buffer layers will heat up due to heat diffusion from FLG into those layers until they have reached their maximum temperature. For bpe, this is when $T_{bpe} = T_{FLG}$, while for BL this is when the heat flux from FLG to BL equals the heat flux from BL to SiC. After a layer reaches its maximum temperature, it will cool down until it reaches the base temperature of 83 K. However, the details of the heating and cooling processes depend strongly on the relative magnitudes of the TBRs.

Figure 3 D shows the results of a simulation with TBRs of $R_K(bpe\leftrightarrow FLG) = 2\cdot 10^{-8}$ K m² W⁻¹, $R_K(FLG\leftrightarrow BL) = 6\cdot 10^{-8}$ K m² W⁻¹ and $R_K(BL\leftrightarrow SiC) = 4\cdot 10^{-8}$ K m² W⁻¹. As indicated, the *T-t* curves can be divided in three distinct regimes: I, II and III. The relative heat flow between the layers in these three regimes is schematically shown in Figure 3 C. In the stage I, the FLG

layer starts at an elevated temperature and thus heat flows quickly to bpe and BL. Consequently, bpe and BL are heating up with characteristic rates determined by $R_K$(bpe↔FLG) and $R_K$(FLG↔BL), respectively; these differences are not resolved in the experiments. In stage III, after the inflection point of the BL curve, bpe, BL and FLG are all cooling down at similar rate. This is in good agreement with the experiments, where $t_{bpe,long} \approx t_{BL}$. This slow overall cooling rate is determined by the specific combination of $R_K$(FLG↔BL) and $R_K$(BL↔SiC). The latter is the most important, since the BL/SiC TBR directly limits the heat flow from the system to the heat sink. The former is important since it affects the temperature of BL, and thus the temperature difference between BL and SiC.

In stage II, different behaviors can be obtained, depending on the relative TBRs. With the chosen set of parameters, the bpe temperature peaks before that of BL and the experimentally observed bi-exponential cooling can be reproduced. As illustrated in Figure 3 C & D, bpe is cooling down throughout regime II. Its cooling rate is faster than in regime III since it closely follows that of FLG. Critically, since FLG can still efficiently lose heat to BL in regime II, the FLG cooling rate is faster than in regime III. Stage II lasts until the inflection point in the BL cooling curve. The time constant for the fast component in the cooling of the bpe layer is determined largely by $R_K$(bpe↔FLG) and $R_K$(FLG↔BL).

In contrast to the bpe curve, the model shows that the BL temperature peaks in the middle of regime II. Additionally, since the BL peak is expected to be much broader, there is a relatively long transition where the BL cooling rate is lower than the final value. Here, the agreement between the simple model system and the experimental data is not perfect; in the experiment the BL temperature reaches its maximum significantly faster and the expected inflection point is not visible. In a separate pump-probe measurement on a clean surface (see **Figure S3**), the inflection is more clearly visible. However, in this measurement the initial rise in temperature is also faster than predicted. This suggests that, contrary to the assumption in the model, BL does absorb some fraction of the incoming pump laser light. Indeed, we have found that a higher initial BL temperature somewhat alleviates the discrepancy. However, a higher initial BL temperature also diminishes the bi-exponential nature of the bpe cooling process. Since the discrepancy in the BL curve cannot be fully resolved within the simple model, the results that exemplify the behavior of bpe more strongly are shown.

In general, good agreement between the simulation and experiment can be obtained for TBR values in the ranges $R_K(\text{bpe}\leftrightarrow\text{FLG}) = 2\pm1\cdot10^{-8}$ K m² W⁻¹, $R_K(\text{FLG}\leftrightarrow\text{BL}) = 6\pm1\cdot10^{-8}$ K m² W⁻¹ and $R_K(\text{BL}\leftrightarrow\text{SiC}) = 4\pm1\cdot10^{-8}$ K m² W⁻¹. Also, a better fit is obtained for low bpe heat capacity values in the order of $C_{p,\text{bpe}} = 300\pm200$ J kg⁻¹ K⁻¹, and high BL heat capacity values in the order of $C_{p,\text{BL}} = 1600\pm300$ J kg⁻¹ K⁻¹. Outside these ranges, a significantly different cooling rate and/or single-exponential cooling of the bpe layer is predicted.

*2.2.3. Discussion*

The TBR resistance values obtained here should be compared to existing literature values. Since understanding the fine details of heat conduction at nanoscopic length scales is of significant importance for many applications,[15-17] TBRs between many technologically relevant materials have been investigated in great detail experimentally and theoretically.[12,23] In general, the magnitude of the TBR can vary several orders of magnitude between different systems. In systems comprising dielectrics and semiconductors, phonons are predominantly responsible for the heat transport and heat conduction through interfaces will depend on phonon scattering and transmission probabilities at that interface. This, in turn, will depend on the temperature, the phonon dispersion relations and the interfacial bonding.[16] Thus, the TBR not only depends on the material properties of the interfacing materials, but also the detailed nature of the interface, including defect density, roughness, and interaction strength. The last factor means, in part, that heat conduction is more efficient in covalently bound systems rather than Van der Waals bound systems.

The TBR values for the FLG-BL and BL-SiC interfaces of $6\pm1\cdot10^{-8}$ K m$^2$ W$^{-1}$ and $4\pm1\cdot10^{-8}$ K m$^2$ W$^{-1}$, respectively, are in good agreement with the literature. In most experiments, TBR values on the order of 0.5 to $5\cdot10^{-8}$ K m$^2$ W$^{-1}$ have been reported for FLG-SiO$_2$[34] and HOPG-Al[35] interfaces. On the other hand, orders of magnitude higher resistances have also been reported for FLG-SiO$_2$[36] and FLG-SiC.[37] Though, in the latter case, the authors attribute the anomalous thermal resistance to interface separation due to significant thermal expansion mismatch between graphene and SiC. Theoretical work, predominantly by means of Molecular Dynamics (MD) simulations, on the TBR of FLG-solid interfaces typically yield TBR values between 0.1 and $10\cdot10^{-8}$ K m$^2$ W$^{-1}$ [38,39] and between 1 and $6\cdot10^{-8}$ K m$^2$ W$^{-1}$ for FLG-SiC interfaces specifically.[40,41] However, it should be noted that these values are either for graphene directly on SiC without a buffer layer or for the total FLG-BL-SiC junction; separate values for FLG-BL and BL-SiC are not usually reported. Also, while the FLG-BL interface might be structurally closer to graphene-graphene interfaces, for which lower resistance values between 0.1 and $1\cdot10^{-8}$ K m$^2$ W$^{-1}$ are found,[38,42] the resistance between FLG and BL is not expected to be that low, since their different structures will lead to mismatched phonon spectra.

A value of $2\pm1\cdot10^{-8}$ K m$^2$ W$^{-1}$ for the TBR of the bpe-FLG interface lies within the expected range. In literature, interfaces between molecules and solid materials in different types of systems have been investigated, both experimentally and theoretically. For solid surfaces or nanoparticles, often modified with a SAM, and molecules in the liquid state, typical TBRs values between 0.1 and $10\cdot10^{-8}$ K m$^2$ W$^{-1}$ are reported.[43] For these systems, the magnitude of the TBR depends mainly on the liquid-solid interaction strength, as given by the hydrophobicity and structure of the surface, and the density of the liquid in the first few MLs. Additionally, MD simulations on Van der Waals bonded FLG-molecular solid interfaces have yielded values between 1 to $5 \cdot 10^{-8}$ K m$^2$ W$^{-1}$.[44] Experimental investigation of the TBR between a solid surface and molecules in a single monolayer is more problematic, since the number of molecules is small.[45] A successful way to gain some insight, and to work towards tailoring interface properties, is the investigation of a molecular monolayer sandwiched between to solid bodies.[46,47] However, in these systems the contribution of a single interface is not often deconvoluted.[46] Still, some results on the thermal barrier between a gold film and SAMs have been reported using ultrafast flash-heating molecular thermal conductance.[22,48] With this technique, values of about $0.45 \cdot 10^{-8}$ K m$^2$ W$^{-1}$ for the TBR between gold and a covalently bound thiol SAM have been reported. The $2 \pm 1 \cdot 10^{-8}$ K m$^2$ W$^{-1}$ we report here falls on the low end of the range of values typically reported for physisorbed molecular layers. This shows that, while not as efficient as chemisorbed SAMs on Au, the thermal coupling between bpe and graphene is relatively strong.

## 3. Conclusion

The bpe/graphene system has been thoroughly characterized with STM and LEED. The observed structure is consistent with the structure reported for bpe on recompressed exfoliated HOPG.[28] Additionally, the structural model was refined by determining the angle between the bpe unit cell and the surface unit cell as 28.4±0.1°. Also, it has been found that the domains of the molecules are not commensurate and on the order of 20 nm in diameter. The desorption temperature of bpe from graphene in UHV was found to be around 300 K. This low desorption temperature prevents the observation of the phase transition that is reported for bpe on recompressed exfoliated HOPG.[28]

Measurements of ultrafast temperature changes and heat dissipation simulations on the bpe/graphene system give an important insight into heat conduction between molecules and the graphene substrate. It was found that the overall heat dissipation from the bpe layer and the buffer layer occurs on a very similar time scale, which is governed by the heat flow from BL to the substrate. However, subtle differences in the early part of the cooling process of these two layers gives an insight into the thermal coupling between the bpe molecules and the graphene layer specifically. It was found that, despite the weak Van der Waals interactions between molecules and surfaces, the heat transfer is still relatively efficient.

The results presented here also show the capabilities of ULEED for the investigation of non-equilibrium heat transfer processes. In future experiments, improved time resolution might enable the possibility of resolving the temperature increase of the layers, which could improve the accuracy in the measurement of the TBRs, since the initial rate of the initial temperature rise depends primarily on the thermal resistance between FLG and the layer in question. Also, in principle a wide range of adsorbates could be investigated, provided they form an ordered layer on graphene, to study the effect of the structural properties of the overlayer on the TBR.

## 4. Experimental Section

*Sample preparation*: FLG was grown on the Si-terminated (0001) face of n-type 6H-SiC by thermal decomposition according to standard procedures.[25] The FLG surface was cleaned in situ by annealing at 400°C for one hour before each deposition. Bpe layers with a sub-ML coverage have been prepared on FLG in two different UHV systems: the STM and the (U)LEED setup. The molecular powder (J&K scientific, 98% pure) was introduced into UHV as received. It was further purified by degassing at RT (STM system) or at 40°C (ULEED system) for several days. Bpe was deposited for 15-30 min with the source kept either at RT[49] (ULEED system) or at approximately 70°C (STM system) and the surface at RT. Immediately after deposition, the sample was cooled using liquid nitrogen to prevent excessive desorption of the bpe molecules.

*STM:* STM measurements were performed in a home-built cryogenic UHV STM system with base pressure better than $5\cdot10^{-11}$ mbar. All STM images were obtained with the sample at 80 K (LNT). STM length scales are calibrated to an external reference and were not corrected for horizontal drift. Large-scale topography images were corrected for vertical drift and tilting by fitting a plane to a physically flat area. Small scale high-resolution images were median filtered (3x3 px), then a line average was subtracted and a plane fit was applied. An idealized image was computed from the raw high-resolution data by tiling of an average unit cell, which was obtained by applying a high-pass filter in Fourier space to remove the long-range contributions and then averaging over 28 unit cells.

*LEED:* (U)LEED measurements were performed in a UHV chamber with base pressure better than $2\cdot10^{-10}$ mbar. The electron source is a home-built laser-driven electron gun as described elsewhere.[9] Electrons are generated from a nanometric tungsten tip by localized two-photon photoemission with a pulsed laser (400 nm central wavelength, 45 fs pulse duration, 20 nJ pulse energy, 25 kHz repetition rate). The LEED patterns were typically recorded at an electron energy of 140-145 eV (clean surface) and 60-65 eV (bpe-covered surface) with a chevron Microchannel Plate, Phosphor Screen, and a cooled sCMOS camera. All LEED patterns were recorded with the sample at 83 K (LNT).

*Time resolved measurements:* ULEED experiments were carried out in an optical-pump/electron-probe scheme. The sample is heated with ultrashort laser pulses (1030 nm central wavelength, 212 fs pulse duration, ~2 mJ cm$^{-2}$ fluence, 25 kHz repetition rate) and probed by an electron pulse of 65 eV energy after a time delay $\Delta t$. The integration time per frame was 30-60 s. No cumulative heating effects were observed. The optimum electron pulse duration at the sample of this setup for an electron energy of 100 eV was determined to be 16 ps.[10] However, at 65 eV temporal broadening of the pulse is increased. Thus, the effective temporal resolution is estimated to be on the order of 40 ps.

To obtain temperature-time (T-t) curves, each frame of the raw pump-probe data was first background-corrected. Then, the intensities of the chosen diffraction spots were summed within a circular area of interest for each time step. The negative logarithm of the intensity was taken to convert the intensity to a value proportional to the temperature change in the system.[32] Potential non-linearities or zero-temperature contributions to the Debye-Waller factor, which might be relevant in 2D systems specifically, were not taken into account.[50] The results were averaged over several diffraction spots in the same pump-probe measurement. A single or bi-exponential decay function was fitted to the *T-t* curves to obtain the characteristic rates.

*Heat diffusion simulations:* Simulations of the temperature evolution in the various layers after the initial heating event were performed with COMSOL Multiphysics® simulation software.[51] For these simulations, a 1D model was adopted, as shown in Figure 3 B. The 1D line segment represents a cut through the material perpendicular to the surface and is divided into four sections corresponding to bpe (1 nm), FLG (2 nm), BL (1 nm), and SiC (2496 nm). Bpe is thermally isolated from the vacuum outside[52] and SiC is connected to an 83 K heat sink. Three TBRs, denoted $R_K$(bpe↔FLG), $R_K$(FLG↔BL) and $R_K$(BL↔SiC), between the sections restrict the heat flux Q across the different boundaries with temperature difference $\Delta T$ according to $Q = \Delta T/R_K$. The temperature evolution of the model system was simulated for 1 ns.

The simulation method is fundamentally a continuum heat conduction model. In general, these types of models break down at small length scales where the assumption of diffusive transport is no longer valid. However, for the description of TBRs specifically, it has been shown that good agreement with more sophisticated models can be obtained if the interfacial resistances are explicitly taken into account.[41,53] Additionally, the 1D geometry neglects heat transport in the surface plane, which can be justified since the size of the experimental laser heating spot is sufficiently large such that the lateral heat diffusion time $t_p \gg 1$ ns. Also, the 1D nature of the model omits any spatial variations in the interfacial thermal resistance between two materials, such as those due to locally varying layer numbers.[54] A model with four sections and three relevant TBRs was used, as it was found that models with fewer TBRs cannot satisfactory reproduce the data.

Finally, an initial state where $T_{0,FLG} = 400$ K and $T_{0,bpe} = T_{0,BL} = T_{0,SiC} = 83$ K was used. Since the parameters that significantly affect the heat flow are temperature independent, this initial temperature is effectively a scaling factor. This initial state does assume that only FLG absorbs the incoming pump laser light. This can be partially justified based on the known optical properties of the materials. Graphene shows significant absorption in this wavelength range[55,56] while absorption by the SiC substrate can be neglected, since the photon energy (1.2 eV) is lower than the bandgap (3.0 eV).[57] Adsorption by bpe cannot be completely excluded, since experimental absorption data is only available for mid-IR and visible frequencies.[58] However, near-IR absorption of molecules is typically quite small.[59] Some absorption by the buffer layer cannot be excluded,[55] but the simulations show that some significant initial temperature difference between FLG and the other layers is required to adequately describe the data.

The primary variables in the model are the three TBRs. However, other material parameters are also relevant to the temperature evolution and thus have to be chosen appropriately. These parameters are the thickness of the material sections, thermal conductivities, densities, and heat capacities. In general, the values for graphite and SiC that are built into the COMSOL Multiphysics® simulation software[52] were used as a baseline for bpe/FLG/BL and SiC, respectively. Since heat conduction is dominated by the TBRs, the results are insensitive to variations in thickness and the thermal conductivities of the layers, which was verified in selected trial simulations. Therefore, the expressions $k_{graphite} = 150$ (W m$^{-1}$ K$^{-1}$)·(300 K T$^{-1}$) and $k_{SiC} = 450$ (W m$^{-1}$ K$^{-1}$)·(300 K T$^{-1}$)$^{0.75}$ were used for the thermal conductivities of bpe/FLG/BL and SiC, respectively. The densities and heat capacities of the layers do have a noticeable effect. Since the effect of the density is the inverse of that of the heat capacity, the densities were fixed at $\rho_{SiC} = 3200$ kg m$^{-3}$ and $\rho_{graphite} = 1950$ kg m$^{-3}$ (FLG and BL), and the density of bpe was estimated to be $\rho_{bpe} = 1162$ kg m$^{-3}$.[28] Finally, all heat capacities were kept temperature independent: the heat capacities of FLG and SiC were set to $C_{p,graphtite} = 710$ J kg$^{-1}$ K$^{-1}$ and $C_{p,SiC} = 1200$ J kg$^{-1}$ K$^{-1}$, respectively, while the heat capacities of bpe and BL were systematically varied to study their effects on the heat dissipation. The influence of the uncertainty in the heat capacities was taken into account when determining the error on the obtained TBRs.

**Supporting Information**
Supporting Information is available from the Wiley Online Library or from the author.

**Acknowledgements**
This work is funded by the European Research Council (ERC Starting Grant 'ULEED', ID: 639119) and the Deutsche Forschungsgemeinschaft (CRC 1073 project C4). We gratefully acknowledge P. Buchsteiner for preparing the graphene substrate, J. G. Horstmann, G. Storeck, and M. Müller for fruitful discussions and the ULEED team for assistance with experiments.

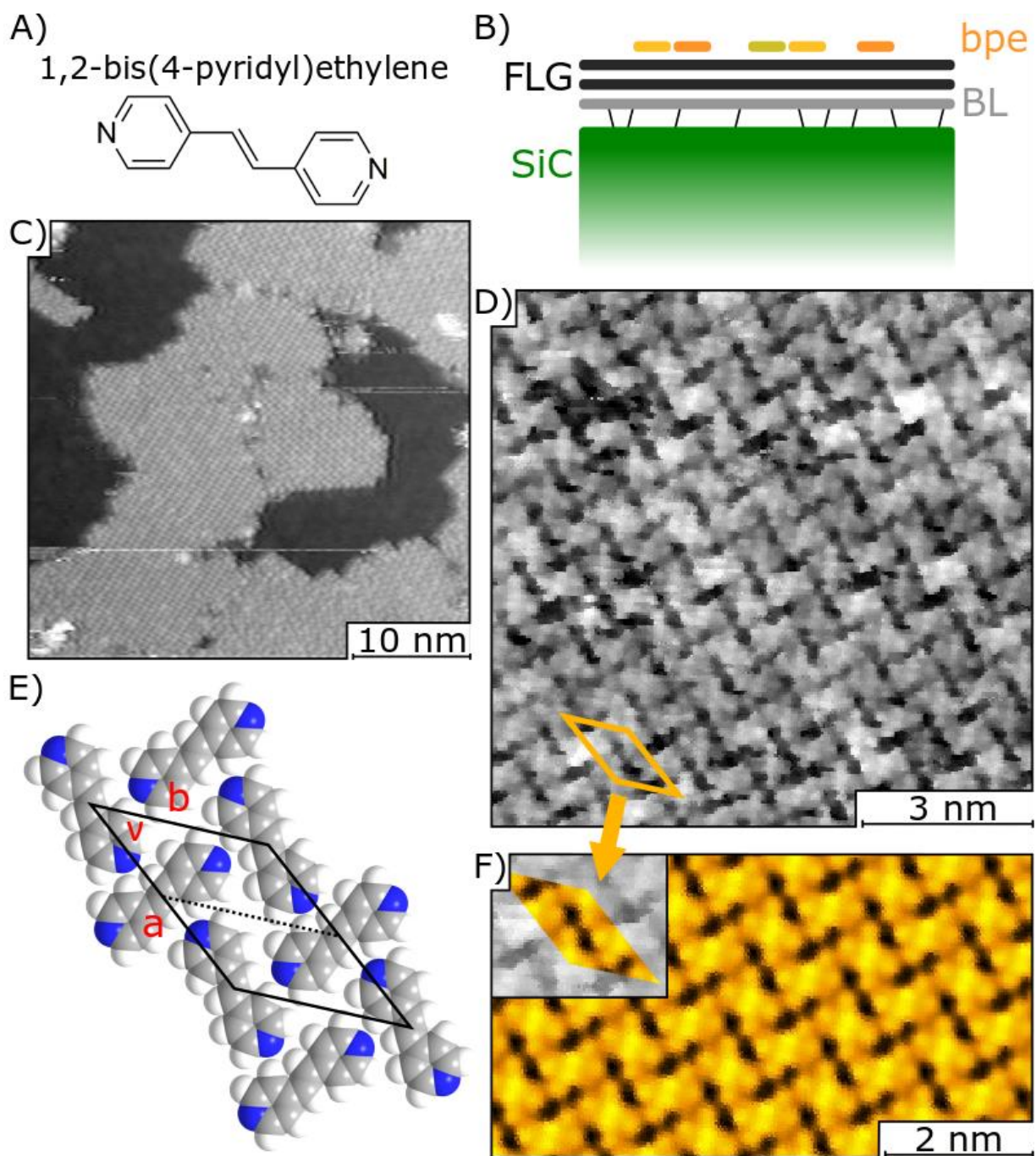

**Figure 1.** A) The structure of bpe. B) The different layers in the system: 6H-SiC bulk, BL and FLG comprise the substrate on which islands of bpe have been grown. C) Large scale overview STM image (1.8 V, 150 pA) of bpe/FLG showing islands with multiple domains. D) A zoomed in STM image (2.1 V, 50 pA) of a single domain. A unit cell is indicated in orange. E) A schematic representation of the unit cell of a bpe island in FLG. The unit cell parameters are: $a$ = 1.777 nm, $b$ = 1.369 nm and $v$ = 39.7° and there are two molecules per unit cell.[28] F) An idealized image, generated by tiling an average unit cell (inset) extracted from D.

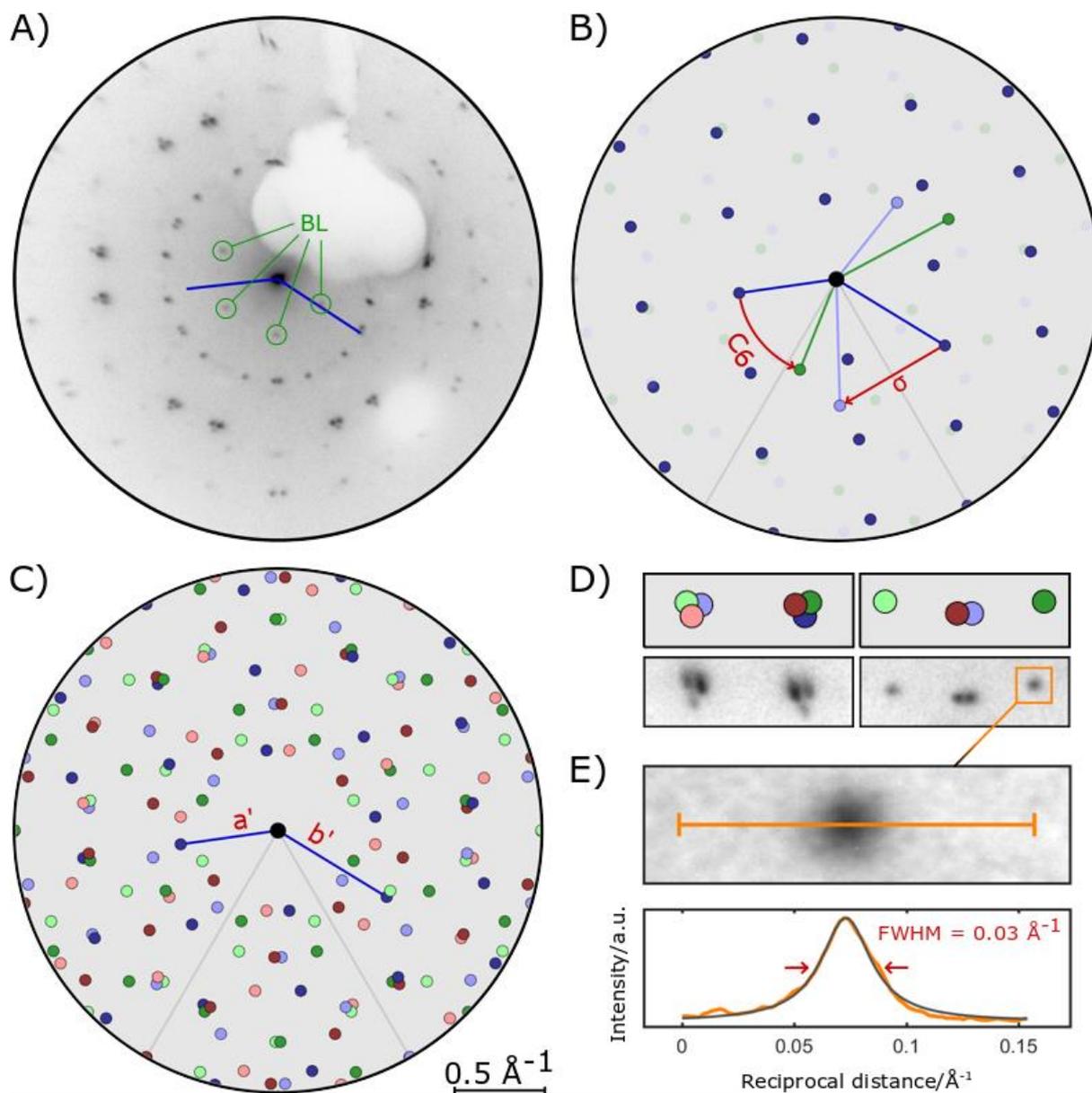

**Figure 2.** A) Experimental LEED pattern of bpe on FLG (LNT, 60 eV, 60 s integration time at 25 kHz repetition rate, plotted with inverse contrast on a log scale). Spots from the (6√3 x 6√3)R30° reconstructed BL (green circles) have been indicated. B) Simulated LEED patterns of a single domain of bpe on FLG. Other domains can be formed by rotating ($C_6$) or mirroring ($\sigma$) this domain as indicated (red arrows). C) Simulated LEED pattern of all six domains of bpe on FLG. Domains with the same brightness (dark or light) are rotational domains whereas domains with the same base color (blue, red or green) are mirror domains. Solid lines in all three diffraction patterns represent reciprocal unit cell vectors of bpe (blue, light blue and green) and of FLG (grey). D) A comparison of two recognizable details that are present in both experimental and simulated patterns. E) A line scan through the indicated diffraction spot (orange line and curve). A fit with a Lorentzian function yields a FWHM of 0.03 Å$^{-1}$ (grey curve).

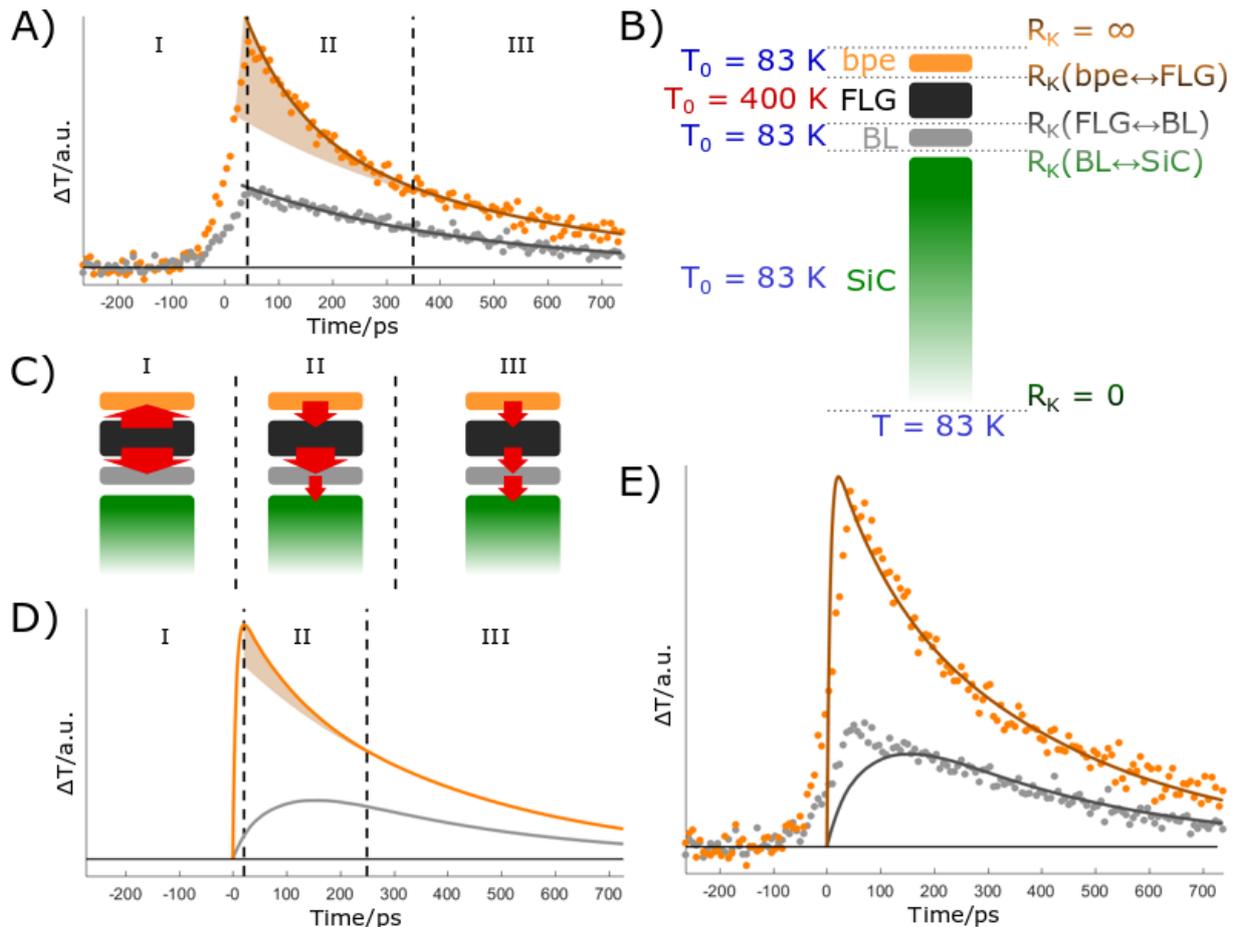

**Figure 3.** A) The temperature evolution of the bpe (orange dots) and BL (grey dots) diffraction peaks as a function of pump-probe delay. A single and a double exponential function has been fitted to the cooling of the two layers respectively (solid lines). The difference between the double exponential function fitted to the bpe data and its slower component has been indicated by the shaded area. B) A schematic illustration of the simulated system, indicating the different layers, the applicable thermal resistances ($R_K$), and the initial temperatures ($T_0$) of the different layers. C) Schematic representation of the heat flow in the different stages I, II and III, after laser heating. The direction and width of the arrows illustrate the net direction and relative magnitude of the heat flow, respectively. D) The temperature evolution of bpe (orange) and BL (grey) extracted from the simulation with thermal resistances $R_K(\text{bpe}\leftrightarrow\text{FLG}) = 2\cdot10^{-8}$ K m$^2$ W$^{-1}$, $R_K(\text{FLG}\leftrightarrow\text{BL}) = 6\cdot10^{-8}$ K m$^2$ W$^{-1}$ and $R_K(\text{BL}\leftrightarrow\text{SiC}) = 4\cdot10^{-8}$ K m$^2$ W$^{-1}$ and heat capacities $C_{p,\text{bpe}} = 300$ J kg$^{-1}$ K$^{-1}$ and $C_{p,\text{BL}} = 1720$ J kg$^{-1}$ K$^{-1}$. The difference between the double exponential function fitted to the bpe data and its slower component has been indicated by the shaded area. E) Comparison of the experimental data in A (dots) and the simulated curves in C (solid line).

Physisorbed sub-monolayers of 1,2-*bis*(4-pyridyl)ethylene on graphene are thoroughly characterized. Moreover, non-equilibrium heat-dissipation from the molecular layer into the surface is investigated. The structure of the substrate enables simultaneous measurement of the temperature of the molecules and the surface. From this, the thermal boundary resistance between the molecules and graphene can be determined.

**Keyword: Monolayers**

Bareld Wit\*, Ole Bunjes, Martin Wenderoth and Claus Ropers\*

**Structure and Non-Equilibrium Heat-Transfer of a Physisorbed Molecular Layer on Graphene**

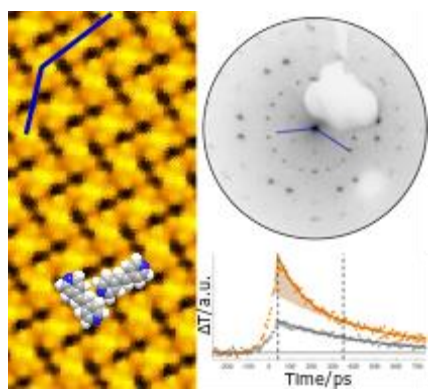

# Supporting Information

**Structure and Non-Equilibrium Heat-Transfer of a Physisorbed Molecular Layer on Graphene**

*Bareld Wit\*, Ole Bunjes, Martin Wenderoth and Claus Ropers\**

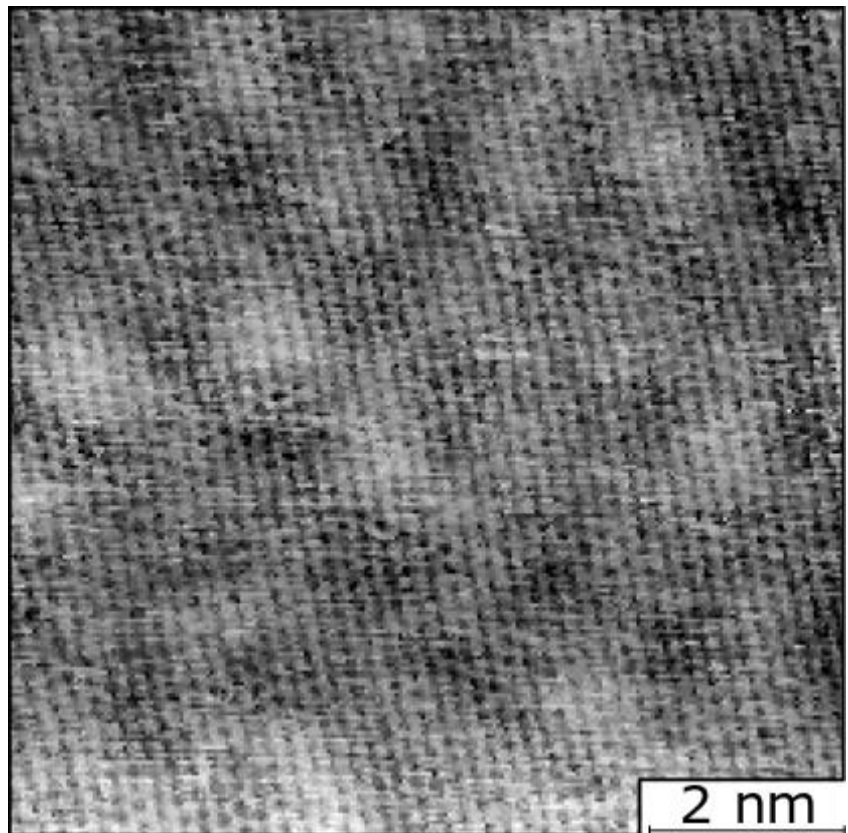

**Figure S1.** STM image (0.05 V, 250 pA) of FLG showing atomic resolution on the graphene as well as the long-range contrast modulation due to the influence of the underlying BL. The image was taken on a bpe covered surface by tunneling through the molecular layer.

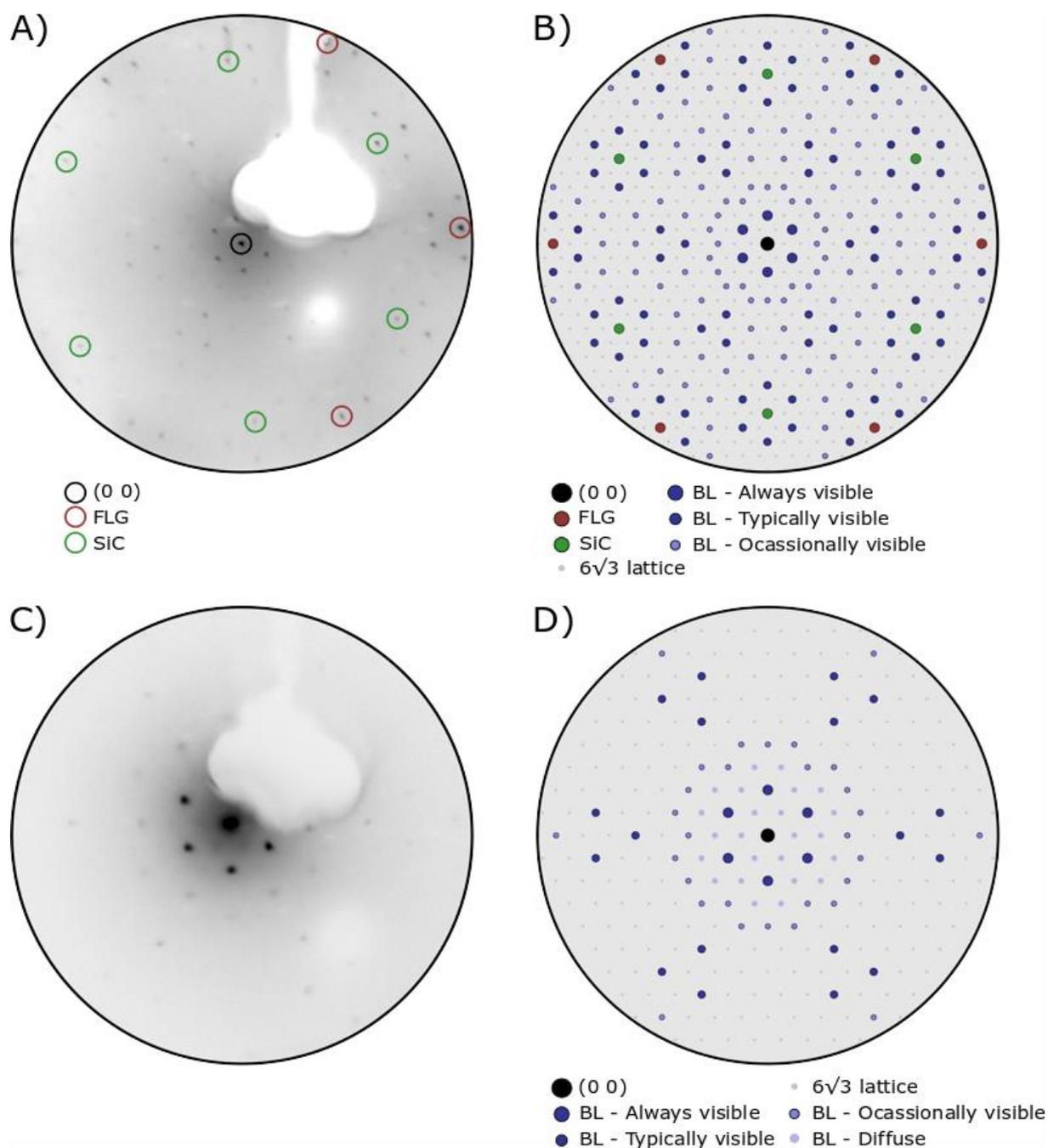

**Figure S2.** A) LEED pattern of clean FLG on SiC (RT, 145 eV, 90 s integration time at 25 kHz repetition rate, plotted with inverse contrast on a log scale). The (0 0) spot and spots originating from FLG and SiC have been indicated. All other spots are due to the (6√3 x 6√3)R30° reconstructed BL. B) Simulated LEED patterns of clean FLG on SiC. The (0 0), FLG and SiC spots have been indicated. Only some of the spots on the (6√3 x 6√3)R30° reconstructed lattice can be observed, as indicated by the differently shaded and sized spots in the simulated pattern. The six spots around (0 0), marked always visible, are observed even on a molecule covered surface (see Figure 2 A). Spots marked typically visible can be observed for the clean surface at most energies. Finally, the spots marked occasionally visible are close to the detection limit, but can be observed after sufficient integration time on a clean surface. C) LEED pattern of clean FLG on SiC (105 K, 60 eV, 180 s integration time at 25 kHz repetition rate, plotted with inverse contrast on a log scale). D) Simulated LEED patterns of clean FLG on SiC. The designation of the spots is the same as in B, with the addition of some diffuse spots near (0 0) that can be observed at this energy.

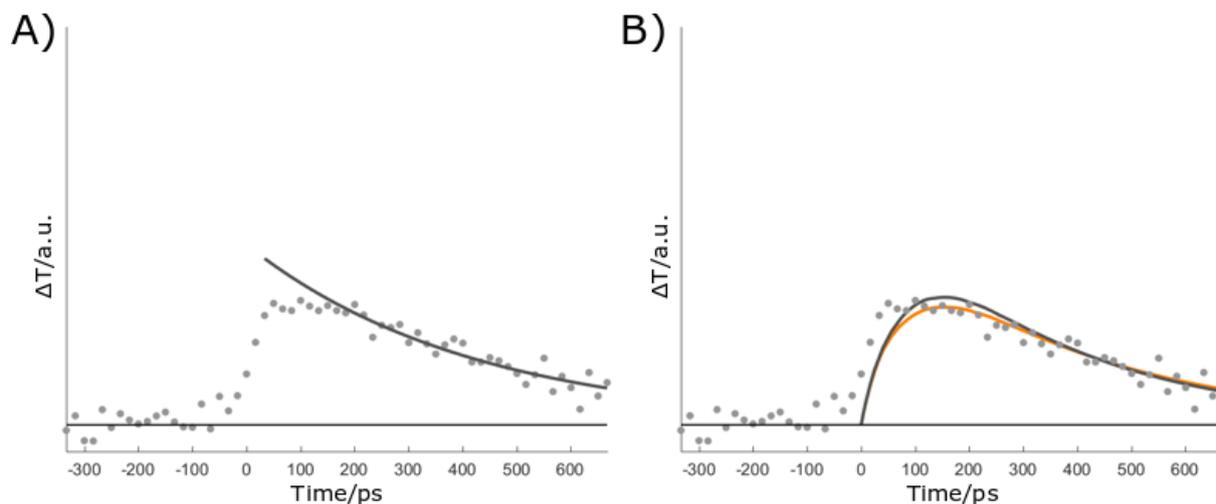

**Figure S3.** A) The temperature evolution of the BL (grey dots) diffraction peaks as a function of pump-probe delay on a clean FLG on SiC sample. A single exponential function has been fitted to the cooling of the layer (solid line). The initial plateau was excluded from the fitting range. B) Comparison of the experimental data in A (dots) and simulated curves (solid lines). The simulated curves are extracted from the model with the same parameters as in Figure 3 ($R_K$(bpe↔FLG) = $2 \cdot 10^{-8}$ K m$^2$ W$^{-1}$, $R_K$(FLG↔BL) = $6 \cdot 10^{-8}$ K m$^2$ W$^{-1}$ and $R_K$(BL↔SiC) = $4 \cdot 10^{-8}$ K m$^2$ W$^{-1}$ and heat capacities $C_{p,bpe}$ = 300 J kg$^{-1}$ K$^{-1}$ and $C_{p,BL}$ = 1720 J kg$^{-1}$ K$^{-1}$). The orange curve represents the case with bpe present, as in Figure 3, and the grey curve represents the case without bpe present.